\documentclass[prl,twocolumn,showpacs,preprintnumbers,amsmath,amssymb]{revtex4}
\usepackage{graphicx}
\usepackage{latexsym}
\usepackage{color}

\begin{document}
\title{Lagrangian temperature, velocity and local heat flux measurement in Rayleigh-B\'enard convection}

\author{Y. Gasteuil, W.L. Shew, M.Gibert, F. Chill\'a, B. Castaing and J.-F. Pinton}
\affiliation{Laboratoire de Physique, de l'\'Ecole Normale
Sup\'erieure de Lyon, CNRS UMR5672, 46 All\'ee d'Italie, 69007 Lyon, France}

\begin{abstract}
We have developed a small, neutrally buoyant, wireless temperature
sensor.  Using a camera for optical tracking, we obtain
simultaneous measurements of position and temperature of the
sensor as it is carried along by the flow in Rayleigh-B\'enard
convection, at $Ra \sim 10^{10}$.  We report on statistics of
temperature, velocity, and heat transport in turbulent thermal
convection. The motion of the sensor particle exhibits dynamics
close to that of Lagrangian tracers in hydrodynamic turbulence. We
also quantify heat transport in plumes, revealing self-similarity
and extreme variations from plume to plume.
\end{abstract}
\maketitle

Understanding fluid motion and transport of heat due to thermal
convection is crucial for progress in diverse challenging and
important problems such as climate change, processes in planetary
and stellar cores, and efficient temperature control in buildings.
Numerous laboratory studies employing Rayleigh-B\'enard
experiments have uncovered laws which relate global heat flux and
flow velocities to fluid properties, flow boundary geometry, and
driving parameters \cite{ConvectionBible}. More recently, several
studies~\cite{Zhou,zhou2} have focused on coherent flow structures, called plumes,
which are ejected from the thermal boundary layers carrying heat
into the convective flow.  We present
measurements from a novel temperature sensor which is carried
along with the convective flow, i.e. Lagrangian measurements.  In
this way we obtain statistics of simultaneous velocity,
temperature, and heat transport dynamics throughout the lifetimes
of many plumes.

Lagrangian measurements are particularly well suited for studying
flows where  coherent structures  \cite{haller} and mixing
\cite{lagmix} are important, e.g. plumes mixing temperature in
convection.  Recent Lagrangian experiments with
submillimeter-sized passive tracer particles have advanced
understanding of turbulence~\cite{Riso,pinton,bodenschatz,ETH}. In
the context of convection, Lagrangian measurements have long been
performed using meter-sized atmospheric balloons~\cite{ballons} or
ocean floats \cite{lagfloat}.  Here we use a similar strategy,
aided by advances in miniature sensors and communication
devices~\cite{JoeRSI}, to probe temperature and flow properties
at centimeter scales in Rayleigh-B\'enard convection.

\begin{figure*}[ht!]
\centerline{\includegraphics[width=0.8\columnwidth]{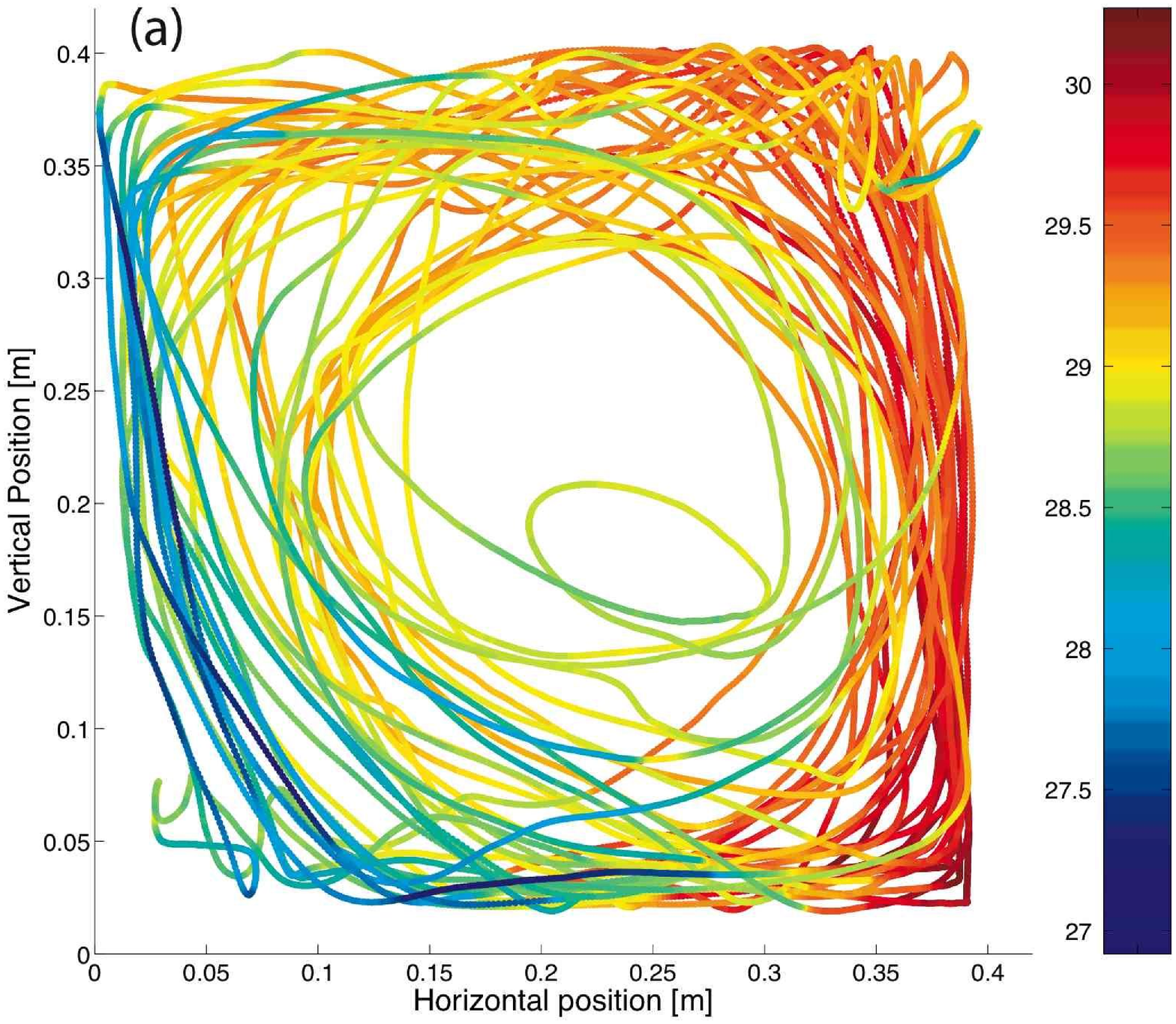}
\includegraphics[width=1.2\columnwidth]{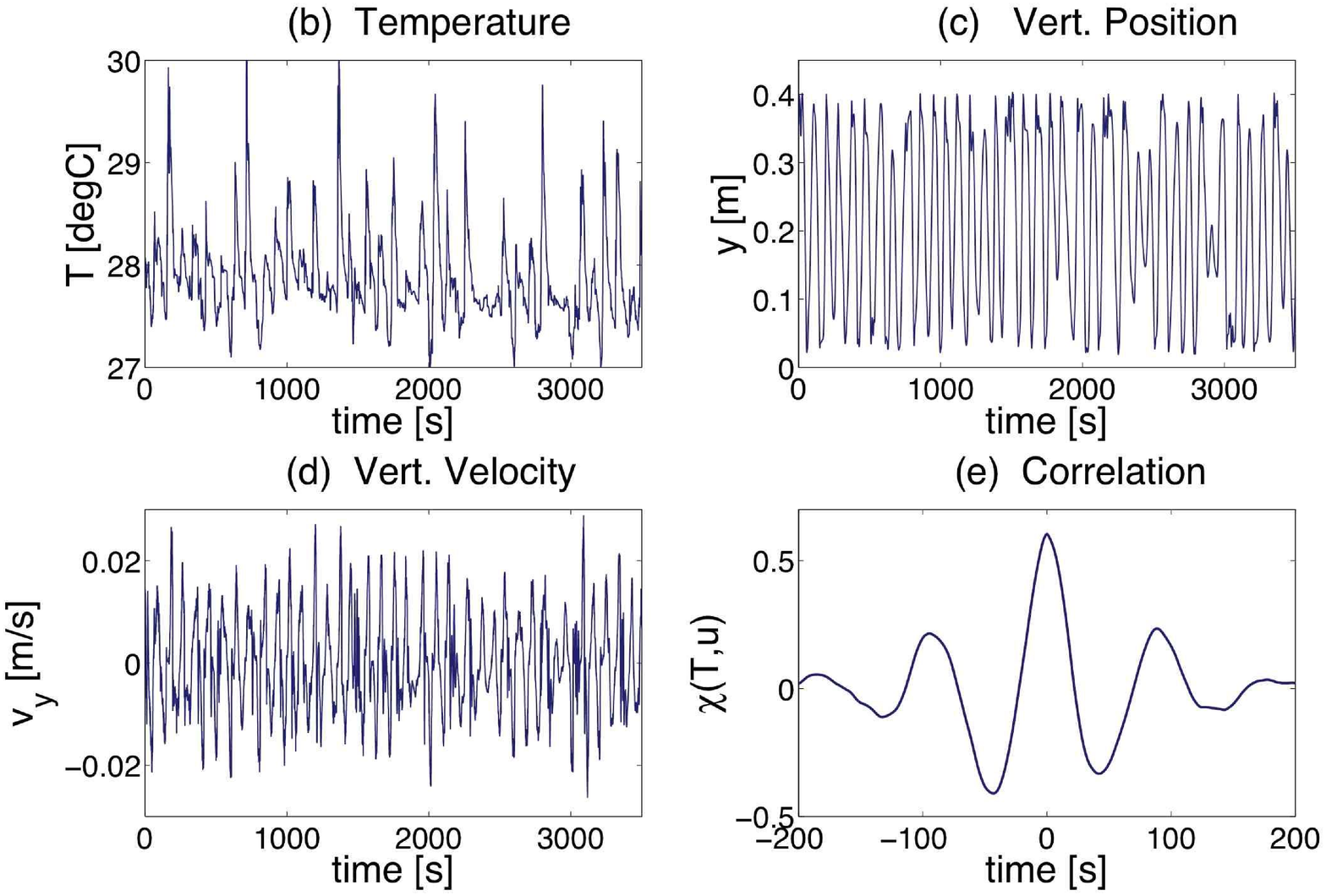}}
\caption{Temperature and velocity measurements of the mobile
sensor. (a) trajectory with temperature color coding; (b)
temperature; (c) vertical position; (d) vertical velocity; (e)
cross correlation of temperature and velocity.} \label{traj}
\end{figure*}

Our experimental setup is a rectangular vessel with height
$H=40$~cm and section $40~$cm~$\times$~10~cm filled with water
(for more detail, see ~\cite{Chilla}).  The walls are made of poly-methylmetacrylate (PMMA), the
top boundary is a copper plate chilled by a controlled water bath,
and the bottom boundary is an electrically heated copper plate.
The heater power is maintained at 230 W and the top plate is held
at $T_{\rm up}=19^\circ$C, resulting in a temperature difference
$\Delta T = 20.3^\circ$C  between the top and bottom plates. The
resulting Rayleigh number is $Ra = {g\beta \Delta T H^3}/{\nu
\kappa} = 3 \times 10^{10}$, where $g$ is acceleration due to
gravity, $\beta = 2.95 \times 10^{-4} \; {\rm K}^{-1}$ is the
thermal expansion coefficient of water and $\nu = 8.17 \times
10^{-7}{\rm m}^{2}{\rm s}^{-1}$, $\kappa = 1.48 \times 10^{-7}{\rm
m}^{2}{\rm s}^{-1}$ its viscosity and thermal diffusivity (values
are given for the mean temperature of the flow 29.1$^\circ$C). The
Nusselt number, defined as the total heat flux normalized by
$\kappa\Delta T/H$, is $Nu = 167.9 \pm 0.2$. Under these
conditions, the convective regime is fully
turbulent~\cite{Chilla,Castaing,Castaing2} and the mean flow is a steady,
system-sized, convection roll with a rotation period of about
100~s.

The mobile sensor consists of a $D = 21$~mm diameter capsule
containing temperature instrumentation, an RF emitter, and a
battery.  It is described in detail in~\cite{JoeRSI} and we
recount its basic features here.  The capsule and fluid density
are carefully matched within 0.05 percent so that it reliably
follows the flow.  Four thermistors (0.8~mm, 230~k$\Omega$,
response time 0.06~s in water) are mounted in the capsule wall
protruding $0.5$~mm into the surrounding flow. A resistance
controlled oscillator is used to create a square wave whose
frequency depends on the temperature of the thermistors. This
square wave is used directly to modulate the amplitude of the
radio wave generated by the RF emitter. The temperature signal is
recovered on-the-fly by a stationary receiver and a Labview
program.  The dynamic range of temperature detection is 80dB, with
a resolution of 4~mK and 50~ms.  In addition, the capsule
trajectory is recorded with a digital video camera, providing
synchronous measurements of the position and temperature of the
sensor as it is carried about by the fluid. With maximum flow
velocities in the range 1-2~cm/s and a particle size of 21~mm, we
are oversampling the dynamics by a factor of order 10. However,
the characteristic thickness of the thermal boundary layer is
$\ell_T \sim \frac{1}{2} H Nu^{-1} \sim 1.2$~mm and that of the
hydrodynamic boundary layer may be estimated~\cite{ConvectionBible} 
as $\ell_U \sim \ell_T(\nu/\kappa)^{+1/3} \sim 2$~mm.  Thus, the sensor is too
large to penetrate the boundary layers.

Before presenting more detailed statistics, we provide an overview
of the raw data collected by the mobile sensor. Predominantly, the
sensor moves in looping trajectories the size of the convection
cell with a period of about 100~s.  The collective result of many
fluid parcels (plumes) with such trajectories compose the large
scale convection roll -- one may also detect in Fig.1a the
presence of secondary rolls in the lower left and upper right
corners. Although the main motion of the sensor is rather
periodic, we point out that its temperature fluctuates widely and
irregularly (see Fig.1b,~c,~d.)  Fig.1e shows the
cross-correlation between temperature and normalized velocity. We
find a maximum value of 0.6 at zero-time lag and confirmation of
the 100~s period of the large scale roll. This correlation is
about twice the value reported in~\cite{Xia2004a} for local
Eulerian measurements.  We attribute these dynamics to the
entrainment of the sensor by thermal plumes and associate the
fluctuations in Figs1.b-d with variation between plumes.  In
addition to qualitative verification of this idea using Schlieren
visualizations, it is consistent with previous studies which find
plumes predominately near the side walls like the trajectory of
our sensor~\cite{Xia2003,Xia2004a}. We explore these dynamics
quantitatively in the results to follow.

From the data in Fig.1, one may estimate characteristic dimensionless numbers for the flow.  From integral quantities one computes the integral Reynolds number $Re=UH/\nu \sim 4000$. The associated turbulent -- Taylor based -- Reynolds number   is of the order of $R_\lambda \propto \sqrt{Re} \sim 63$. We note that this value is very close to a local Reynolds number, defined from the actual motion of the Lagrangian sensor $Re'= u_{rms} \ell_{rms} /\nu \sim 60$, where $\ell_{rms}$ is the fluctuation of position along a `mean' trajectory, $u_{rms}$ is the usual standard deviation of the velocity.  In this regime ($R_\lambda \sim 60$), one expects turbulence with significant intermittency in the velocity gradients~\cite{LathropNature}.

Accordingly, we turn now to more detailed statistics of our
measurements including power spectra, increment probability
distribution functions (PDF), and structure function scaling for
velocity and temperature.  The power spectra (in time) are shown
in Fig.2a.  The velocity spectrum is close to the behavior
expected for a Lagrangian tracer in a turbulent flow -- $f^{-2}$
-- although scaling range is limited at such low $R_\lambda$. The
temperature spectra is observed to roughly mimic that of velocity
with a steeper power law slope compared to measurements with a
stationary temperature probe.  For such Eulerian measurements,
experiments~\cite{7over5} measure slopes between -1.35  and -1.4 ($\equiv -7/5$)  while 
numerical studies~\cite{Verzicco2004a} report -7/5 in the bulk and -5/3
near the side walls. Velocity increments (Fig.2b) indicate that
the sensor is subjected to rather intermittent acceleration with
strongly non-Gaussian statistics -- flatness is -21.  Like the
power spectrum, the shape of PDFs for different velocity increments
is similar to those of a fluid particle in a turbulent flow -- Fig.2b.
Finally, we have computed the evolution of the structure functions for
the vertical velocity $S^p_{vz}(\tau) = \langle | v_z(t+\tau) -
v_z(t) |^p \rangle_t$; we observe an extended region of relative
scaling $S^p_{vz}(\tau) \propto [ S^2_{vz}(\tau) ]^{\xi_p}$ (for
$0.3 \; {\rm s} \leq \tau \leq 10 \; {\rm s}$) with exponents
$\xi_1=0.56, \xi_2=1.00, \xi_3=1.30, \xi_4 = 1.50, \xi_5=1.65,
\xi_6=1.7$. These values (determined with a 10\% precision) 
are again in good agreement with experimental
measurements for Lagrangian tracers in turbulence~\cite{pinton}. We also show in Fig.2c the probability density function (PDF) of temperature increments measured by the mobile sensor. Similar to Eulerian measurements~\cite{Xia2004a}, their
statistics are non Gaussian, with wider tails at small scale. In
contrast to the velocity behavior (and to some Eulerian
studies~\cite{Verzicco2004a}), our Lagrangian temperature
increments do not reveal a range of self-similar scaling.

\begin{figure}[th]
\centerline{\includegraphics[width=1.1 \columnwidth]{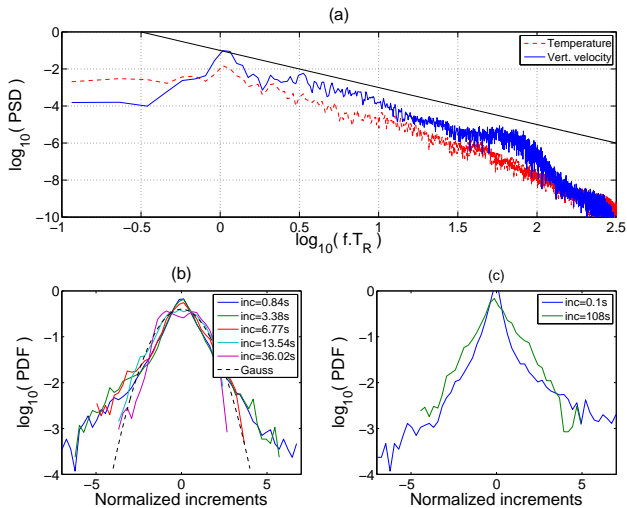}}
\caption{Temperature and velocity statistics. (a) Time spectra of
temperature and vertical velocity of the sensor particle; the
frequency is non-dimensionalized using the period $T_R = 100$~s of
the large scale roll motion and curves have been shifted
vertically for clarity). The solid straight line corresponds to an ideal $f^{-2}$ scaling; (b) PDFs of vertical velocity increments (time lags are given in the inset); (c) PDFs of temperature increments for time lags equal to 0.1 and 102.4~s.} \label{Tinc}
\end{figure}

As proposed in several previous studies~\cite{Xia2003,Xia2004b,Lohse2004}, the dynamics of heat transport may be best analyzed in terms of a local heat flux
$v_z \theta'(t)$, where $\theta'(t) = \theta(t) - \overline{\theta}$ is the particle
temperature variation from its time averaged temperature $\overline{\theta}$, and $v_z(t)$ its vertical velocity. Our mobile sensor allows for a Lagrangian measurement of this quantity. In particular, we define a normalized Lagrangian vertical heat transport
\begin{equation}
Nu^{L}(t) = 1 + \frac{H}{\kappa \Delta T} \; v_z(t) \theta'(t) \ .
\end{equation}
The time series $Nu^L(t)$ is shown in Fig.3a. As expected, it is most often
positive since convective motions are associated with either hot
fluid rising or cold fluid sinking, in each case $( v_z\theta') > 0$.  The much less probable events with $( v_z\theta') < 0$ correspond, for instance, to the rise of the particle when it
is colder than its environment due to turbulent swirls in the
flow. We find that the time averaged Lagrangian heat transport
$\overline{Nu^L}=328$ is larger than the global value
($Nu=168$ in our case).  This indicates that our sensor
preferentially samples the regions of the flow which carry higher
than average heat, i.e. plumes.  In contrast, the traditional
Nusselt number indiscriminately accounts for all regions in the
flow such as the center of the cell where plumes rarely visit.
Another prominent feature is the highly non-Gaussian, intermittent
fluctuations of Lagrangian heat transport, which is similar to
measurements with stationary probes at similar Rayleigh
numbers~\cite{Xia2004a}.

\begin{figure}[th]
\centerline{\includegraphics[width=1.0\columnwidth]{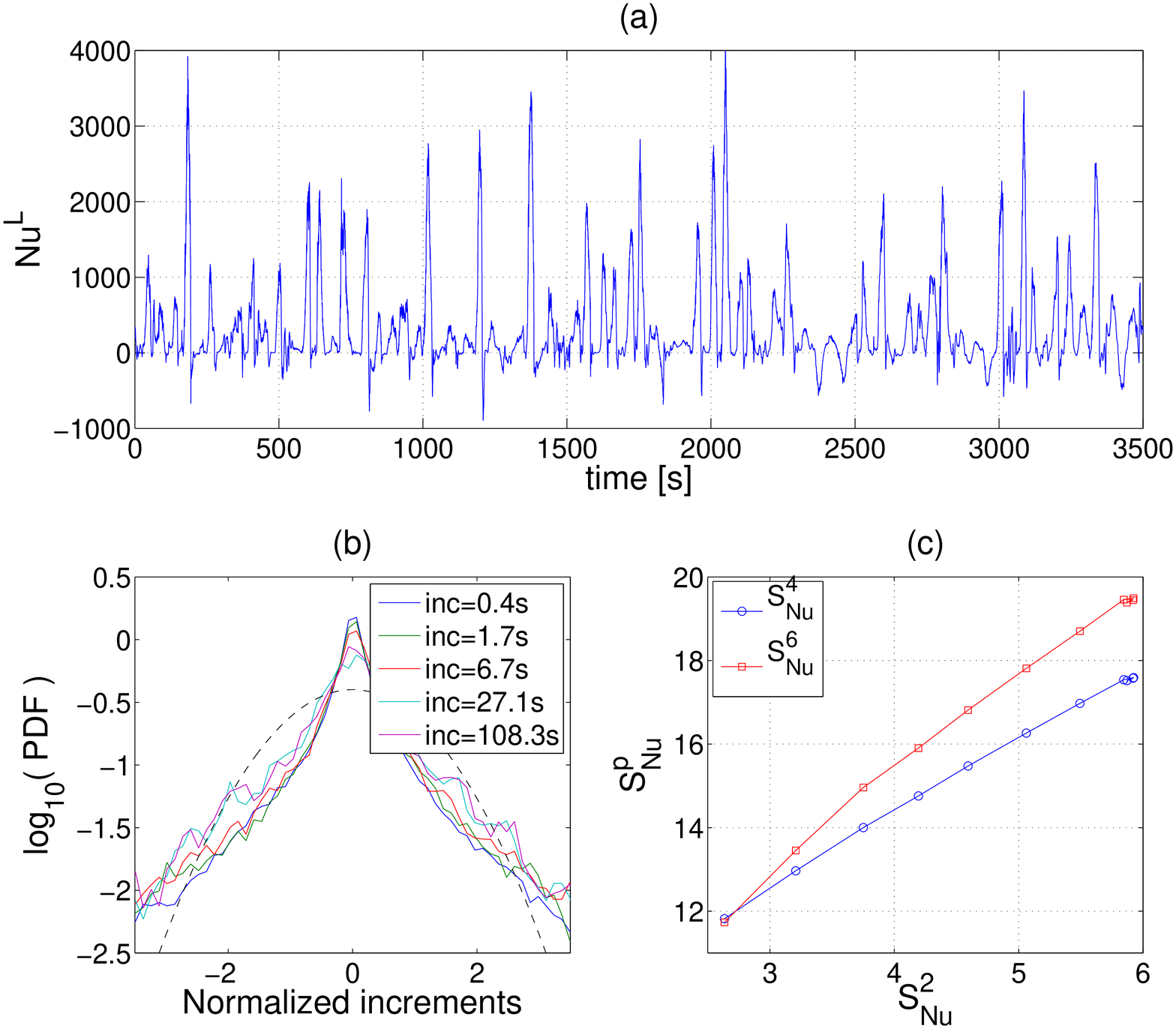}}
\caption{Lagrangian heat transport. (a) time series; (b) PDF of
increments, and Gauss distribution
(dashed line). (c) ESS plot of  fourth and sixth order structure
functions, {\it vs} the second order one (the second order
structure function has been shifted vertically).} \label{Nuinc}
\end{figure}

To further quantify Lagrangian heat transport, we have studied the
statistics of the increments (in time) $Nu^L(t+\tau) - Nu^L(t)$.
Two features are noteworthy: (i) as shown in Fig.3b, their PDFs
are strongly non Gaussian from the smallest time increments to
lags of the order of integral time $T_R$; (ii) there exists a
range of relative scaling for their structure functions
$S^p_{Nu}(\tau) = \langle | Nu^L(t+\tau) - Nu^L(t) |^p \rangle_t$, as
shown in Fig.3c. The relative exponents, $S^p_{Nu}(\tau) \propto [
S^2_{Nu}(\tau) ]^{\xi^{Nu}_p}$ have values $\xi^{Nu}_1=0.54,
\xi^{Nu}_2=1.00, \xi^{Nu}_3=1.37, \xi^{Nu}_4 = 1.68,
\xi^{Nu}_5=1.93, \xi^{Nu}_6=2.14$ -- the use of the second order
structure function as a reference is arbitrary. 
Comparison with the velocity shows that the Lagrangian heat flux
is less intermittent; i.e. the PDFs of the local heat flux in
Fig.3b are non Gaussian but their shapes do not evolve
significantly as increments increase towards the large scale
period $T_R$.

A simple interpretation of the above statistical results is that thermal plumes may be defined as Lagrangian heat transport events (once detached from the thermal boundary layer); these events are in some way self-similar as will become clear below. In this light, we have analyzed the portions of the sensor
trajectories during which its vertical position lies between $H/4$
and $3H/4$, i.e. the rise (or fall) of a hot (cold) plume. For
each of these motions, we compute the mean heat transport, and its
fluctuation measured by the standard deviation computed over the
portion of the trajectory. The results are reported in Fig.4.
We observe that the heat transport has large variations from one
plume to the next.  The mean Lagrangian heat transport in these events  ($\overline{Nu^{\rm pl}} = 335$) is quite close to the mean
for the entire time series ($\overline{Nu^L} = 328$), which
indicates that the times the sensor is not in a plume are
relatively unimportant to heat transport. Furthermore, the
variation from plume to plume is quite large, ${\rm rms}(Nu^{\rm
pl}) = 384$, of the order as the mean.  During the trajectory of a
single plume, the standard deviation $\sigma$ of $Nu^{\rm pl}$ is
proportional to the mean; $\sigma = \alpha \overline{Nu^{\rm pl}}$
where $\alpha=0.5 \pm 0.05$.  We note that the plume heat
transport we measure matches within a factor of 2 the prediction
of the model proposed in~\cite{Lohse2004}, though more experiments
are required to test scaling laws for different Rayleigh numbers and fluid
properties. Furthermore, we find that the characteristic time for
temperature decay during the life-time of a plume is of the order
of its travel time from one plate to the other, which also
supports the model in~\cite{Lohse2004}.

\begin{figure}[h]
\centerline{\includegraphics[width=1.0\columnwidth]{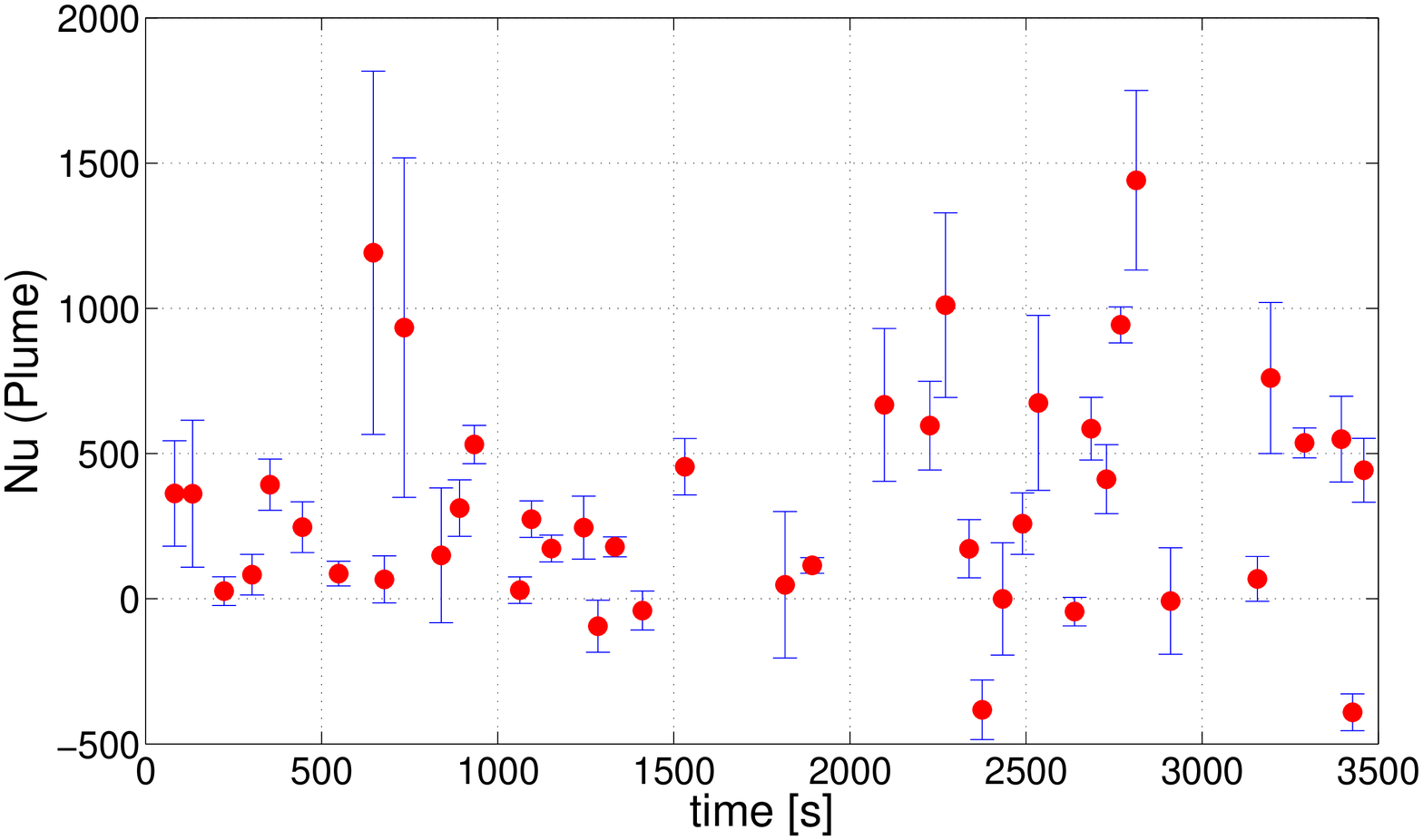}}
\caption{Plumes. Mean (dots) and standard deviation (error bars)
of Lagrangian heat transport of consecutive plumes. }
\label{EventFluct}
\end{figure}


To summarize, we have reported novel measurements of Lagrangian temperature and heat
transport using a wireless, neutrally buoyant temperature sensor.
The sensor provides a new perspective on the dynamics of thermal
plumes in turbulent Rayleigh-B\'enard convection.  We find that
heat transport fluctuates greatly from one plume to another and
that these fluctuations suggest a self-similar character of
plumes.  A practical implication of our results is that in order
to maximize heat transport, one should maximize the production of
thermal plumes. This is supported by global $Nu$ versus $Ra$
measurements in experiments~\cite{cilibe1,Verzicco2004} with rough endplates.
Future investigations will be focused on the scaling of plume heat
transport for varying $Ra$ and different fluids.\\

\noindent {\bf Acknowledgements} \\
This work has been supported by Emergence Rh\^one-Alpes Contract
No. 2005-12 and CNRS.  We acknowledge useful discussions with
Sergio Ciliberto and Romain Volk.






\end{document}